\documentclass[apjl]{emulateapj}
\usepackage{graphicx}

\newcommand{\kms}{\,km\,s$^{-1}$}
\newcommand{\spd}{{\sc spd}}
\newcommand{\Teff}{\ensuremath{T_{\rm eff}}}                
\newcommand{\Msun}{\ensuremath{\,{\rm M}_\odot}}            

\renewcommand{\ion}[2]{{#1}\,{\sc {\small{#2}}}}
\newcommand{\er}[3]{\ensuremath{#1^{+#2}_{-#3}}}
\newcommand{\Apx}{\,\AA\,px$^{-1}$}                                     

\shorttitle{Spin-orbit misaligment in AS\,Cam?}
\shortauthors{Pavlovski, Southworth and Kolbas}

\begin{document}

\title{Apsidal motion of the eclipsing binary AS Camelopardalis: discrepancy resolved}

\author{
K. Pavlovski\altaffilmark{1,2},
J. Southworth\altaffilmark{2}
and V. Kolbas\altaffilmark{1}
}
\altaffiltext{1}{Department of Physics, Faculty of Sciences, University of Zagreb, Zagreb, Croatia}
\altaffiltext{2}{Astrophysics Group, Keele University, Staffordshire, ST5 5BG, UK}





\begin{abstract}
We present a spectroscopic study of the eclipsing binary system AS\,Camelopardalis, 
the first such study based on phase-resolved CCD \'echelle spectra. Via a spectral
 disentangling analysis we measure the minimum masses of the stars to be 
$M_{\rm{A}}\sin^3i=3.213\pm0.007$\Msun\ and $M_{\rm{B}}\sin^3i=2.323\pm0.006$\Msun, 
their effective temperatures to be $\Teff({\rm{A}})=12\,840\pm120$\,K and 
$\Teff({\rm{B}})=10\,580\pm240$\,K, and their projected rotational velocities to be
 $v_{\rm{A}}\sin{i_{\rm{A}}}=14.5\pm0.1$\kms\ and $v_{\rm{B}}\sin{i_{\rm{B}}}\leqslant4.6\pm0.1$\kms.
 These projected rotational velocities appear to be much lower than the synchronous values. 
We show that measurements of the apsidal motion of the system suffer from a degeneracy between
 orbital eccentricity and apsidal motion rate. We use our spectroscopically-measured $e=0.164\pm0.001$
 to break this degeneracy and measure $\dot\omega_{\rm{obs}}=0.133\pm0.010^\circ$\,yr$^{-1}$. 
Subtracting the relativistic contribution of $\dot\omega_{\rm{GR}}=0.0963\pm0.0002$$^\circ$\,yr$^{-1}$
 yields the contribution due to tidal torques: $\dot\omega_{\rm{cl}}=0.037\pm0.010$$^\circ$\,yr$^{-1}$.
 This value is much smaller than the rate predicted by stellar theory, 0.40--0.87$^\circ$\,yr$^{-1}$.
We interpret this as a misalignment between the orbital axis of the close binary and the rotational 
axes of its component stars, which also explains their apparently low rotational velocities. 
The observed and predicted apsidal motion rates could be brought into agreement if the stars 
were rotating three times faster than synchronous about axes perpendicular to the orbital axis. 
Measurement of the Rossiter-McLaughlin effect can be used to confirm this interpretation.
\end{abstract}



\section{Introduction}                                                                                                               \label{sec:intro}

Tidal torques dominate the dynamical evolution of close binary systems. In order of expected timescales, tidal effects should firstly align the stellar rotation axes with the orbital axis, synchronize the stellar rotational to the orbital frequency, and then circularize the orbit \citep{Mazeh08eas}. These effects happen on timescales of millions to billions of years, depending sensitively on the characteristics of individual systems. On a much shorter timescale apsidal motion, the precession of an eccentric orbit in its own plane, can produce an observable rate of change in the longitude of periastron, $\dot\omega=d\omega/dt$.

AS\,Camelopardalis is a detached eclipsing binary containing two late-B stars with an orbital period of 3.43\,d and a disputed apsidal motion period of order 2000\,yr. The apsidal period, $U$, is much longer than expected for the measured properties of AS\,Cam, leading in the past to concerns about our understanding of stellar physics and even of general relativity \citep{Maloney89aj,Maloney91aj}. The apsidal motion of AS\,Cam was discovered by Khaliullin \& Kozyreva (1983) from eclipse timings, and found to have a period of $U=2250\pm200$\,yr. \citet{Maloney89aj} obtained a similar $U=\er{2400}{630}{1300}$\,yr. However, \citet{Krzesinski90ibvs} and \citet{BozkurtDegirmenci07mn} found very different apsidal periods, $920\pm470$\,yr and $740\pm6$\,yr respectively, by adopting lower eccentricity values of $e\approx0.10$. These shorter apsidal periods are closer to theoretical expectations, but the low $e$ conflicts with other observations \citep{Hilditch72pasp,Maloney91aj}. In addition, a third body has been found orbiting AS\,Cam with a period of $P_3=2.2$\,yr \citep{KozyrevaKhaliullin99arep,BozkurtDegirmenci07mn}, albeit with a low statistical significance.

The binaries AS\,Cam and DI\,Herculis have for many years been the two best-known systems with apsidal periods much longer than they should be \citep{GuinanMaloney85aj}. In a breakthrough work \citet{Albrecht+09natur} observed the Rossiter-McLaughlin effect in DI\,Her and showed that its orbital and rotational axes are misaligned. This lengthens the expected apsidal period, which is now within 10\% of the observed value \citep{Claret10aa}. Here we present the first high-resolution time-resolved spectroscopy of AS\,Cam, from which we find that the projected rotational velocities of the stars are much lower than expected and that their rotational axes are likely misaligned with the orbital axis. Below we refer to the primary (hotter and more massive) star as star\,A, and to the secondary as star\,B.


\section{Observations and data reduction}                                                                                              \label{sec:obs}

\begin{figure*} \centering \includegraphics[width=0.9\textwidth]{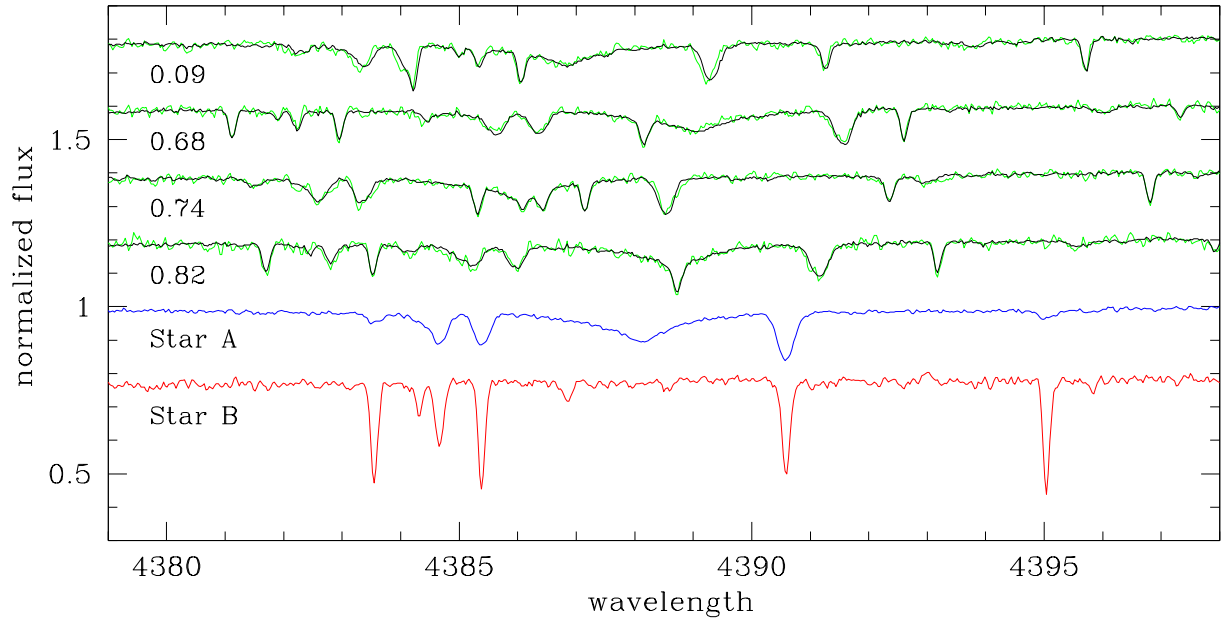}
\caption{\label{fig:4385} Example spectra of AS\,Cam in the region of the \ion{He}{i}
4388\,\AA\ line, offset from unity by arbitrary amounts for display purposes. The top
four spectra (green lines) are observed composite spectra, with the orbital phases of
observation annotated. The lower two spectra are the disentangled spectra for star\,A
(blue line) and star\,B (red line). The disentangled spectra have been recombined and
overplotted on the observed spectra (black lines) to show the quality of the fits.}
\end{figure*}

We obtained 31 high-resolution spectra of AS\,Cam over five nights in 2007 October, using the Nordic Optical Telescope and high-resolution Fibre-fed Echelle Spectrograph. Wavelength scales were established from thorium-argon exposures taken regularly each night. We opted for medium resolving power ($R=48\,000$) by using fibre 3 in bundle B. This gave complete spectra coverage in the interval 3640--7360\,\AA\ at a reciprocal dispersion ranging from 0.023\Apx\ in the blue to 0.045\Apx\ in the red, at a resolution of approximately 3.5\,px. An exposure time of 600\,s was used for all spectra, resulting in continuum signal-to-noise (S/N) ratios of roughly 90--140 per pixel.

Five additional spectra were secured in service mode on 2010/10/10, in this case in the high-resolution ($R=67\,000$) mode with fibre 4 in bundle B. The spectral format was identical to the medium-resolution mode but the resolution was 2.4\,px. Exposure times of 900\,s gave S/N ratios of about 250 per pixel.

Basic reduction of these data (bias subtraction, flat-fielding, scattered light correction, extraction of orders, and wavelength calibration) was performed with IRAF. Removal of the instrumental blaze function was not trivial because the broad Balmer lines of B-stars can extend over entire \'echelle orders. In such cases we interpolated between blaze functions from adjacent orders, using a semi-manual approach and {\sc{java}} routines written by VK. This process had to be undertaken carefully, as spectral disentangling is very sensitive to the normalization of the observed spectra.


\section{Spectral disentangling analysis}                                                                                              \label{sec:spd}

We subejcted our observations to a spectral disentangling (\spd) analysis to determine the masses and atmospheric properties of the components of AS\,Cam. The \spd\ approach \citet{SimonSturm94aa} allows the best-fitting (in a least-squares sense) orbital elements and separated spectra of a binary star system to be obtained directly from a set of observed composite spectra covering a range of orbital phases (Fig.\,\ref{fig:4385}). The resulting disentangled spectra have a very high S/N ratio as they contain the total signal in the input spectra (in this case about 440 for star\,A and 180 for star\,B), so are well suited to further analysis such as chemical abundance determination \citep{PavlovskiHensberge05aa,PavlovskiMe09mn,Pavlovski+09mn}. \spd\ does not require template spectra and is not biased by any blending of the spectral lines of the two stars in the observed spectra, so is excellent for measuring orbital elements \citep[e.g.][]{MeClausen07aa}. A review of the theoretical and practical aspects of \spd\ can be found in \citet{PavlovskiHensberge09xxx}.

We disentangled the spectra of AS\,Cam in Fourier space using the {\sc{FDbinary}}\footnote{\tt http://sail.zpf.fer.hr/fdbinary/} code \citep{Ilijic+04aspc}, concentrating on 10--15 relatively narrow spectral regions (50--80\,\AA\ wide) which contain only metallic lines. Both stars exhibit numerous and very sharp spectral lines, resulting in well-defined orbital elements. Our adopted elements (Table\,\ref{tab:orbit}) are the mean and standard error of the results from the individual spectral regions. An important result is that eccentricity $e=0.164\pm0.001$, which conclusively rules out those apsidal motion studies which found short apsidal periods and $e\approx0.10$ (Sect.\,\ref{sec:intro}). The velocity amplitudes of \citet{Hilditch72pasp}, obtained using photographic methods, are in reasonable agreement with our values.

\begin{table*}\centering
\caption{\label{tab:orbit}The orbital elements and atmospheric properties of AS\,Camelopardalis from spectral disentangling.}
\setlength\tabcolsep{25pt}
\begin{tabular}{l r@{\,$\pm$\,}l  r@{\,$\pm$\,}l}\hline
Parameter                                       & \multicolumn{2}{c}{Star\,A} & \multicolumn{2}{c}{Star\,B} \\ \hline
Orbital period (d)                              & \multicolumn{4}{c}{3.430973 (fixed)}                      \\
Time of periastron pasage (HJD)                 & \multicolumn{4}{c}{$2454399.75211 \pm 0.0026$}            \\
Orbital eccentricity $e$                        & \multicolumn{4}{c}{$0.164 \pm 0.001$}                     \\
Longitude of periastron $\omega$ ($^\circ$)     & \multicolumn{4}{c}{$241.45 \pm 0.50$}                     \\
Velocity amplitude (\kms)                       & $106.22$ & $0.14$           & $146.92$ & $0.13$           \\
Mass ratio $q$                                  & \multicolumn{4}{c}{$0.723 \pm 0.002$}                     \\
Minimum mass $M\sin^3i$ (\Msun)                 & $3.213$ & $0.007$           & $2.323$ & $0.006$           \\
Effective temperature $\Teff$ (K)               & $12\,840$ & $120$           & $10\,580$ & $240$           \\
Projected rotational velocity $v\sin{i}$ (\kms) & $14.5$ & $0.1$              & $\leqslant4.6$ & $0.1$      \\
\hline
\ \\
\end{tabular}\end{table*}

\begin{figure}\includegraphics[width=0.48\textwidth]{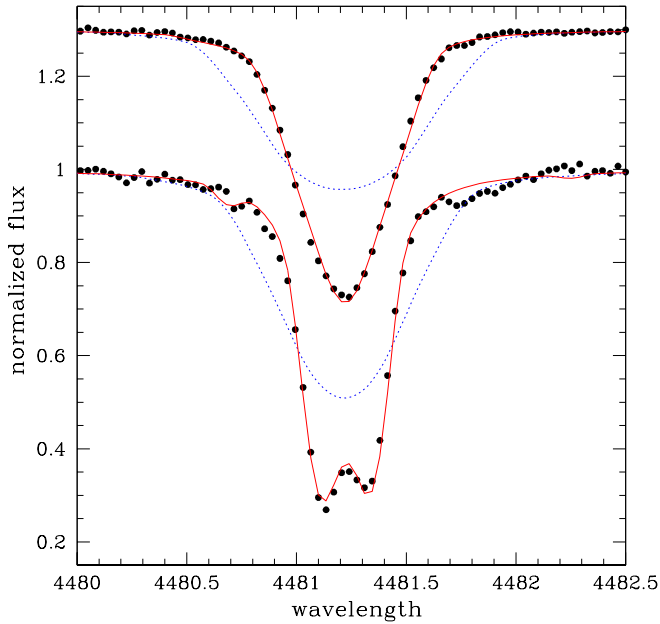}
\caption{\label{fig:4481}Profiles of the \ion{Mg}{ii} 4481\,\AA\ line for
star\,A (offset by $+0.3$) and star\,B (normalized to unity). Synthetic
spectra for the measured $v\sin{i}$ values are shown with red solid lines.
Blue dotted lines show the synthetic spectra broadened to the synchronous
rotational velocities.}\end{figure}

We now turn to the measurement of effective temperature (\Teff) from Balmer line profiles. H$\beta$ and H$\gamma$ profiles for the two stars were obtained by disentangling the observed spectra with the orbital elements fixed at the values in Table\,\ref{tab:orbit}. The disentangled spectra are not normalized to the correct continuum because such information is not available in the input spectra (unless some were taken during eclipse; \citealt{Tamajo11aa}). We therefore renormalized the Balmer profiles to the continuum using the light ratios of the two stars obtained by modelling the $UBV$ light curves of AS\,Cam from \citet{Hilditch72mmras}. The $\Teff$ values were then obtained by fitting the profiles with synthetic H$\beta$ and H$\gamma$ spectra calculated using the {\sc{uclsyn}} program \citep{Smalley01,Smith92phd}. We fixed the surface gravities of the stars to known values, $\log{g_{\rm{A}}}=4.154\pm0.013$ and $\log{g_{\rm{B}}}= 4.278\pm0.014$ (Southworth et al., in preparation). The blue and red sides of each Balmer profile were fitted separately, resulting in four \Teff\ measurements for each star. Our final values are the mean and standard deviations of these: $\Teff({\rm{A}})=12\,840\pm120$\,K and $\Teff({\rm{B}})=10\,580\pm240$\,K.

The projected rotational velocities of the components were measured using a set of isolated metal lines, mostly of \ion{Fe}{ii}, \ion{Ti}{ii} and \ion{Cr}{ii}, in the 4400-5000\,\AA\ region. The instrumental broadening was obtained for each \'echelle order from the thorium-argon spectra. Representative values are $6.5\pm0.1$ and $4.8\pm0.1$\kms\ for the 2007 and 2010 data, respectively. Each line was then fitted with {\sc{uclsyn}} synthetic spectra, yielding $v_{\rm{A}}\sin{i_{\rm{A}}}=14.5\pm0.1$\kms\ and $v_{\rm{B}}\sin{i_{\rm{B}}}\leqslant4.6\pm0.1$\kms. These are much lower than the synchronous velocities, which are $36.7\pm0.5$\kms\ and $27.1\pm0.5$\kms, respectively. In contrast, \citet{Hilditch72pasp} found that the $v\sin{i}$ values were approximately synchronous. We attribute this discrepancy to the long exposure times (roughly 1\,hour) needed to obtain photographic spectra of AS\,Cam, which will have caused smearing of the spectral lines due to orbital motion. Fig.\,\ref{fig:4481} shows the \ion{Mg}{ii} 4481\,\AA\ line profiles for the two stars compared to synthetic spectra calculated for the measured and for synchronous $v\sin{i}$ values.


\section{Apsidal motion}                                                                                                            \label{sec:apsmot}

Published measurements of the apsidal period and eccentricity of AS\,Cam are conspicuous by their disagreemt. Below we cast the argument in terms of the apsidal motion rate in degrees per year. $\dot\omega_{\rm{obs}}$ is the observed rate, $\dot\omega_{\rm{GR}}$ is the relativistic contribution and $\dot\omega_{\rm{cl}}$ is the classical (tidal) contribution which can be estimated from stellar theory.


\citet{KhaliullinKozyreva83apss} discovered apsidal motion in the AS\,Cam system, finding $e=0.1695\pm0.0014$ and $\dot\omega_{\rm{obs}}=0.16^\circ$\,yr$^{-1}$. \citet{Maloney89aj} adopted $e=0.17$ and got $\dot\omega_{\rm{obs}}=0.150\pm0.053$$^\circ$\,yr$^{-1}$, noting that this was far smaller than the expected amount due to $\dot\omega_{\rm{GR}}=0.085\pm0.002$$^\circ$\,yr$^{-1}$ and $\dot\omega_{\rm{cl}}=0.358\pm0.058$$^\circ$\,yr$^{-1}$. \citet{Krzesinski90ibvs} suggested that a lower eccentricity could at least partially solve this problem, obtaining $e=0.10\pm0.01$ and $\dot\omega_{\rm{obs}}=\er{0.39}{0.80}{0.26}$$^\circ$\,yr$^{-1}$. \citet{Wolf96aas} adopted $e=0.14$ and found $\dot\omega_{\rm{obs}}=0.183\pm0.026$$^\circ$\,yr$^{-1}$, whereas \citet{KozyrevaKhaliullin99arep} assumed $e=0.17$ to obtain $\dot\omega_{\rm{obs}}=0.149\pm0.015$$^\circ$\,yr$^{-1}$. Finally, \citet{BozkurtDegirmenci07mn} arrived at $e=0.1018\pm0.0006$ and $\dot\omega_{\rm{obs}}=0.486\pm0.004$$^\circ$\,yr$^{-1}$. This $\dot\omega_{\rm{obs}}$ is consistent with $\dot\omega_{\rm{GR}}+\dot\omega_{\rm{cl}}$, although its errorbar is certainly too small.

\begin{figure} \centering \includegraphics[width=0.48\textwidth]{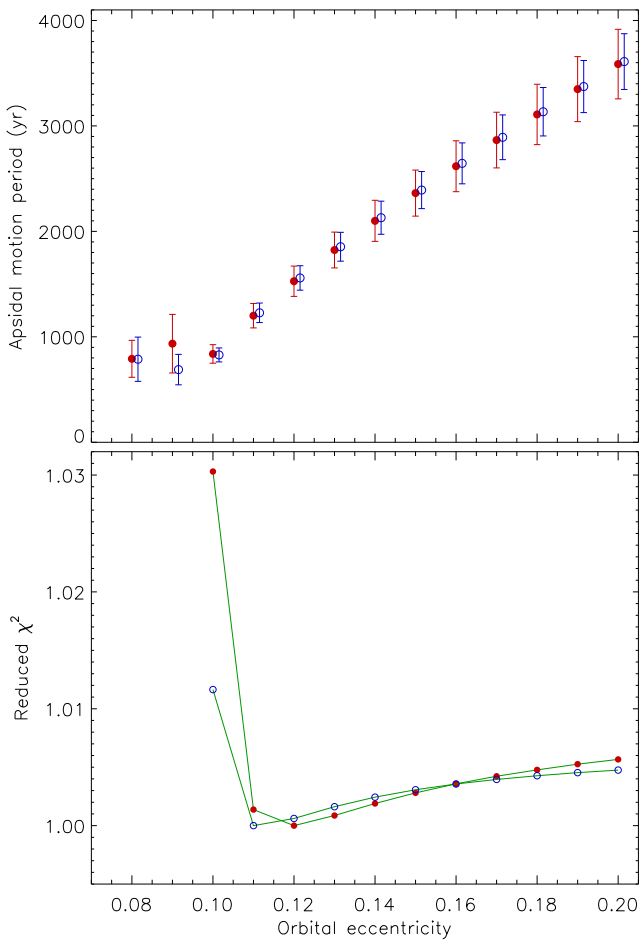}
\caption{\label{fig:apsmot}Plot of the apsidal motion periods found for AS\,Cam
as a function of eccentricity (upper panel). Filled red and open blue circles
show results respectively with and without including photographic observations.
The lower panel shows $\chi^2_\nu$, which is forced to equal $1.0$ for the best
fit in each case.} \end{figure}

From the above details it is clear that $e$ and $\dot\omega_{\rm{obs}}$ are highly correlated for AS\,Cam, whose apsidal period is much longer than its observational history. To illustrate this we collected all available times of minimum light for AS\,Cam and analysed them using the {\sc{jktapsmot}} code \citep{Me04mn2}, which implements the exact ephemeris-curve method of \citet{Lacy92aj2}. Most of these minimum timings have no associated errorbar, so uncertainties were estimated to be 0.01\,d for photographic results and 0.001\,d for photoelectric/CCD timings. All errorbars were then scaled up by a factor of five during the fitting process to force $\chi^2_\nu = 1.0$ for the best fit. Solutions were then made both with (124 measurements) and without (92 measurements) the photographic timings, and for eccentricities of from 0.08 to 0.20. The upper panel in Fig.\,\ref{fig:apsmot} demonstrates the strong correlation between $e$ and $\dot\omega_{\rm{obs}}$, whereas the lower panel shows quality of the fit. Whilst the lowest $\chi^2_\nu$ is found for $e\approx0.115$, all solutions in the range $e=0.11$--$0.20$ are statistically acceptable.

We have a crucial advantage over previous studies, namely the very precise $e=0.164\pm0.001$ found from our \spd\ analysis (Table\,\ref{tab:orbit}). We can therefore reject low-eccentricity solutions \citep{Krzesinski90ibvs,BozkurtDegirmenci07mn} with extreme confidence, and also break the degeneracy between $e$ and $\dot\omega_{\rm{obs}}$. Rerunning the {\sc{jktabsdim}} analysis with $e$ fixed to $0.164$ and including all times of minimum, we find $\dot\omega_{\rm{obs}}=0.133\pm0.010^\circ$\,yr$^{-1}$, corresponding to $U=2700\pm250$\,yr.

From the orbital elements of AS\,Cam (Table\,\ref{tab:orbit}) we measure the relativistic contribution to the apsidal motion to be $\dot\omega_{\rm{GR}}=0.0963\pm0.0002$$^\circ$\,yr$^{-1}$ \citep[e.g.][]{Gimenez85apj}. Subtracting this from the observed value gives the rate due to classical effects: $\dot\omega_{\rm{cl}}=0.037\pm0.010$$^\circ$\,yr$^{-1}$.

\begin{figure}\includegraphics[width=0.48\textwidth]{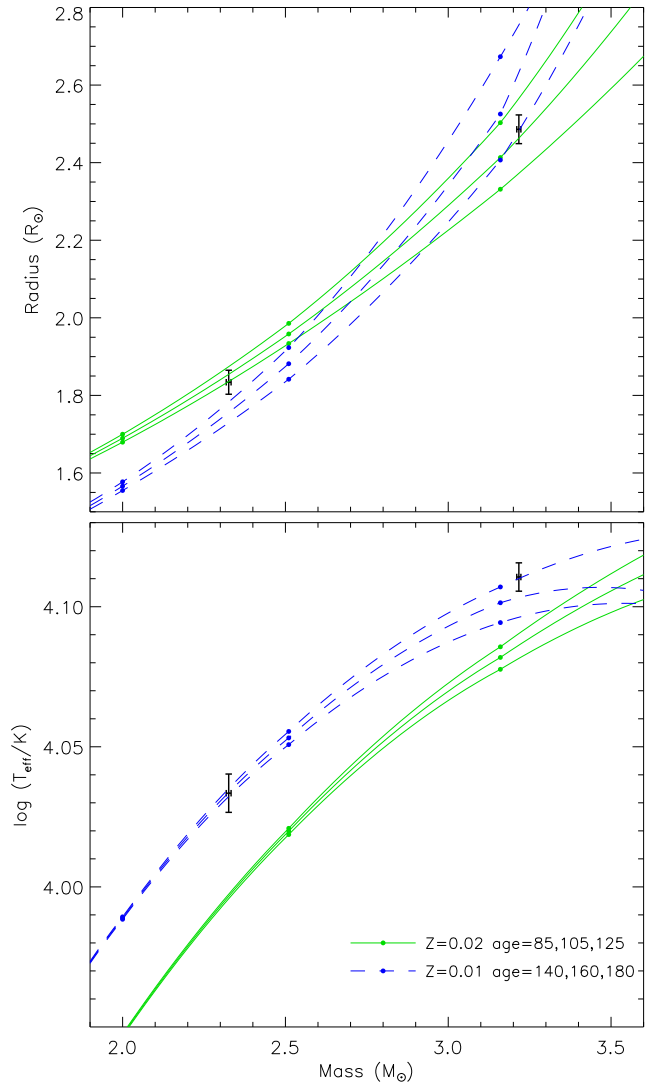}
\caption{\label{fig:evo}Positions of the stars in AS\,Cam in the mass--radius
and mass--\Teff\ diagrams, compared to theoretical predictions from the Granada
models. Results are shown for fractional metal abundances by mass of $Z=0.02$
(green solid lines) and $Z=0.01$ (blue dashed lines) and for the best-matching
ages $\pm$20\,Myr. The filled circles represent the points at which tabulated
model predictions are available, and the lines show a quadratic interpolation
between these points.} \end{figure}

The expected $\dot\omega_{\rm{cl}}$ can be obtained from the internal structure constants $\log{k_2}$, using the equations given by \citet{ClaretGimenez93aa}. In turn, $\log{k_2}$ must be estimated from theoretical stellar structure models, and depends on the detailed characteristics of the stars. We have compared AS\,Cam to the Granada model tabulations \citep{Claret95aas,Claret97aas,ClaretGimenez95aas,ClaretGimenez97aas} using the minimum masses from Table\,\ref{tab:orbit}, orbital inclination and radii from Southworth et al.\ (in preparation), and $\Teff$ values from Sect.\,\ref{sec:spd}. From Fig.\,\ref{fig:evo} we find that the fractional metal abundance of the binary is in the range $Z = 0.01$--0.02, which is qualitatively in good agreement with the spectral line strengths, and that its age is about 130\,Myr.

The internal structure constants for $Z=0.02$ and age 105\,Myr are $\log{k_2}({\rm{A}})=-1.55$ and $\log{k_2}({\rm{B}})=-1.50$, whereas for $Z=0.01$ and age 160\,Myr they are $-2.33$ and $-1.46$. $\log{k_2}({\rm{A}})$ is rather sensitive to the evolutionary status of star\,A. These $\log{k_2}$ values result in theoretical $\dot\omega_{\rm{cl}}$ values of 0.87 and 0.40$^\circ$\,yr$^{-1}$, for $Z=0.02$ and $Z=0.01$ respectively. These are both well in excess of the 0.037$^\circ$\,yr$^{-1}$ which is actually observed. We therefore confirm the discrepant apsidal motion of AS\,Cam to a high level of significance.

The third body orbiting AS\,Cam has a very small light-time amplitude and an orbital period of 2.2\,yr \citep{KozyrevaKhaliullin99arep,BozkurtDegirmenci07mn}, which is orders of magnitude shorter than $U$ so should not have a significant effect on the analysis above. We have checked this by calculating a periodogram of the residuals of the apsidal-motion fit, using the {\sc{Period04}} package \citep{LenzBreger04iaus}. We find a peak of 2$\sigma$ significance at $P_3=824$\,d. This period is consistent with previous studies, as expected because most minimum timings are in common. Our attempts to fit a spectroscopic orbit to the residuals led to a wide variety of solutions, depending on the starting parameter values and which parameters were fitted. New apsidal motion solutions with these orbits subtracted are not significantly different to our baseline solution. We conclude that there may be a third body, but that the data in hand are insufficient to confirm its existence or give its orbital parameters, and that in any case it does not affect our measured $\dot\omega_{\rm{obs}}$.


\section{Prognosis}

We find that AS\,Cam has a classical apsidal motion rate which is an order of magnitude lower than theoretically predicted. How can this be explained?

\subsection{Problems with stellar theory or gravity theory}

The predicted $\dot\omega_{\rm{cl}}$ relies on our understanding of stellar physics, which is certainly imperfect. But it is difficult to see how the rate could decrease by an order of magnitude in order to match our observations. Modern stellar theory does a good job of explaining apsidal motion in the great majority of close binaries \citep{ClaretGimenez10aa} so AS\,Cam would have to be a special case. Similar comments apply to relativistic apsidal motion, as the theory of general relativity is otherwise highly successful in a wide range of scientific disciplines.

\subsection{Anomalously slow rotation}



The $v\sin{i}$ values of the stars are slower than the synchronous values by factors of 2.5 (star\,A) and 6 (star\,B). This apparently slow rotation results in a smaller predicted $\dot\omega_{\rm{cl}}$, but only by $\sim$10\%. Thus slow rotation cannot explain the $\dot\omega_{\rm{cl}}$ discrepancy.

Tidal effects are expected to move the rotation rates towards the synchronous values and $e$ towards zero, so it is reasonable to ask why we see such low $v\sin{i}$ values. The synchronization timescales \citep{Zahn75aa} are 106\,Myr for star\,A and 1300\,Myr for star\,B. As the age of the system is 100--160\,Myr, we don't expect the stars to have reached synchronisation yet. They would thus have had to form with very slow rotation rates, which is exceptional but at least more plausible for a close binary than for a single star \citep{Tohline02araa}. The orbital circularization timescale is much longer again, in agreement with the observed eccentricity.

\subsection{Spin-orbit misalignment}

AS\,Cam and DI\,Her are two well-known binaries with anomalously low $\dot\omega_{\rm{obs}}$ values which challenge our understanding of stellar physics. The problem with DI\,Her has recently been explained as resulting from a large misalignment between the orbital and rotational axes \citep{Albrecht+09natur}. This is a highly plausible explanation for AS\,Cam, particularly due to the low $v\sin{i}$ values we find for both stars. \citet{Maloney91aj} considered this possibility for AS\,Cam itself, but rejected it because the stars were then thought to be rotating synchronously \citep{Hilditch72pasp}.

In the light of the considerably subsynchronous rotation which we find for AS\,Cam, it is interesting to reconsider this hypothesis. \citet{Shakura85sval} found that the line of apsides can undergo retrograde motion if the stellar rotational axes are not aligned to the orbital axis, with the largest effect when the axes are perpendicular.  In the case of AS\,Cam, Shakura found that the observed and computed $\dot\omega_{\rm{cl}}$ values could be brought into agreement if the stars were rotating at three times the synchronous rate around axes perpendicular to the orbital axis. It follows that the stars' rotation axes should be tilted by 82$^\circ$ and 87$^\circ$ with respect to the orbital axis to produce the observed $v\sin{i}$ values, which is indeed very close to perpendicularity.

Although the axial misalignment hypothesis is very persuasive, it does incur the question of why tidal effects have not aligned the axes. The timescale for axial alignment is much shorter than for rotational synchronization \citep{Hut81aa,Mazeh08eas}, and therefore much shorter than the age of AS\,Cam. Possible reasons for a misalignment between spin and orbital axes have been discussed by \citet{Bonnell+92apj}. Axial misalignment may also explain the slow rotational velocities found for the eclipsing binary systems V615\,Persei \citep{Me04mn} and the central star of the planetary nebula SuWt\,2 \citep{Exter+10aj}.

\subsection{Perturbations from a third body}

The times of minimum light of AS\,Cam suggest (but do not require) the presence of a third body causing a light-time effect (Sect.\,\ref{sec:apsmot}). Extensive numerical calculations have shown that a third star in an orbit almost perpendicular to the orbital plane of a close binary can cause anomalous apsidal motion \citep{Khodykin04apj,FabryckyTremaine07apj,Borkovits07aa}. This option seems unlikely, but cannot be dismissed as yet.


\section{Summary}

We have presented the first modern spectroscopic study of AS\,Cam, an eclipsing binary which shows anomalously slow apsidal motion. Through a spectral disentangling approach we have obtained high-precision measurements of the minimum masses of the stars, their \Teff s and their projected rotational velocities. If the stellar rotational axes are aligned with the orbital axis, the stars are rotating much more slowly than the synchronous velocities. We have re-investigated the apsidal motion of the system and demonstrated the strong correlation between apsidal motion rate and eccentricity. We have used our precise measurement of $e=0.164\pm0.001$ to break this degeneracy, finding $\dot\omega_{\rm{obs}}=0.133\pm0.010^\circ$\,yr$^{-1}$, corresponding to $U=2700\pm250$\,yr. The relativistic component of this is $\dot\omega_{\rm{GR}}=0.0963\pm0.0002$$^\circ$\,yr$^{-1}$, so the tidal component is therefore $\dot\omega_{\rm{cl}}=0.037\pm0.010$$^\circ$\,yr$^{-1}$. From theoretical stellar evolutionary models we predict a very different $\dot\omega_{\rm{cl}}$ in the range 0.40--0.87$^\circ$\,yr$^{-1}$.

We find no reason to suspect problems with our understanding of stellar physics or general relativity, primarily because the $\dot\omega_{\rm{obs}}$ values for most other close binaries agree well with theoretical predictions. Invoking slow rotation only changes the predicted $\dot\omega_{\rm{cl}}$ by $\sim$10\% so does not solve the discrepancy. However, the low $v\sin{i}$ values suggest that the rotational axes of the stars are highly inclined with respect to the orbital axes. \citet{Shakura85sval} found that the discrepant apsidal motion for AS\,Cam could be explained if the rotational axes were perpendicular to the orbital axis and that the stars were rotating three times faster than synchronously. We therefore interpret our results as evidence of axial misalignment in the AS\,Cam system. The same phenomenon was found for DI\,Her \citep{Albrecht+09natur}, using a different observational approach.

Observation of the Rossiter-McLaughlin effect in AS\,Camelopardalis would allow further constraints to be placed on its dynamical characteristics \citep[e.g.][]{Albrecht+07aa,Albrecht+09natur,Albrecht+11apj}, as would measuring additional times of minimum light. Further theoretical study of the tidal effects in misaligned binary systems would also be very illuminating.


\section*{Acknowledgments}

KP acknowledges receipt of the Leverhulme Trust Visiting Professorship which made possible his sabbatical stay at Keele University. The work of KP is funded by the Croatian Ministry of Science. JS would like to thank STFC for the award of an Advanced Fellowship. This paper is based on observations made with the Nordic Optical Telescope, operated on the island of La Palma jointly by Denmark, Finland, Iceland, Norway, and Sweden, in the Spanish Observatorio del Roque de los Muchachos.


\bibliographystyle{mn_new}

\end{document}